\documentclass[slac_one]{revtex4}
\usepackage{graphicx}
\usepackage{fancyhdr}
\usepackage{slashed}
\usepackage{amssymb}
\begin{document}

\title{Search for Supersymmetry Using the Trilepton Signature of Chargino-Neutralino Production}

\author{J. Glatzer (for the CDF collaboration)}
\affiliation{Department of Physics and Astronomy, Rutgers University, Piscataway, New Jersey 08854}

\begin{abstract}
We use the three lepton and missing energy ``trilepton'' signature to search for chargino-neutralino
production with $2.0\,\mathrm{fb^{-1}}$ of integrated luminosity collected by the CDF II experiment at the Tevatron
$p\bar{p}$ collider. We expect approximately 11 supersymmetric events for a specific choice of parameters
of the mSUGRA model, but our observation of 7 events is consistent with the standard model
expectation of 6.4 events. We constrain the mSUGRA model of supersymmetry and rule out chargino
masses up to $145\,\mathrm{GeV/c^2}$ for a specific choice of parameters.
\end{abstract}

\maketitle

\thispagestyle{fancy}

\section{Supersymmetry}
Supersymmetry is a hypothetical symmetry between bosons and fermions and predicts a fermion (boson) ``superpartner'' for every standard model boson (fermion). It doubles the number of particles and is able to address several open questions in the standard model. The supersymmetric model mSUGRA is widely used due to its small number of free parameters: the common scalar mass $m_0$, the common gaugino mass $m_{1/2}$, the ratio of the vacuum expectation values of the two Higgs doublets $\tan\beta$, the common trilinear coupling $A_0$ and the sign of $\mu$, where $\mu$ is the Higgs mixing parameter. Charginos $\tilde{\chi}_i^\pm$ and neutralinos $\tilde{\chi}_i^0$ are the mixed mass eigenstates of the superpartners of the gauge and Higgs bosons. 

At the Tevatron associated pairs of $\tilde{\chi}_1^\pm$ and $\tilde{\chi}_2^0$ can be produced in $p\bar{p}$ collisions with reasonable cross section. The leptonic decay of $\tilde{\chi}_1^\pm\to l^\pm \nu_l\tilde{\chi}_1^0$ and $\tilde{\chi}_2^0\to l^+ l^-\tilde{\chi}_1^0$ via off-shell $W^\pm$, $Z^0$ boson or on- or off-shell  sleptons $\tilde{l}_R^\pm$ as intermediate states is a promising search channel at hadron colliders. The signature is three leptons and missing transverse energy from the weakly interacting LSP's $\tilde{\chi}_1^0$ and the neutrino, which are not detectable

\newcommand{\ttt}{l_tl_tl_t}
\newcommand{\ttl}{l_tl_tl_l}
\newcommand{\tll}{l_tl_ll_l}
\newcommand{\ttT}{l_tl_tT}
\newcommand{\tlT}{l_tl_lT}

\section{The Analysis}
We identify electrons and muons; ``loose'' electrons or muons need to satisfy weaker requirements than ``tight'' ones. Isolated tracks are used to indicate the hadronic decay of the $\tau$ lepton to one charged particle. Based on the type of the selected analysis objects we define five analysis channels: channels with three leptons $\ttt$, $\ttl$, $\tll$ and channels with two leptons and an isolated track $\ttT$, $\tlT$, where $l_t$, $l_l$ and $T$ refer to a tight lepton, a loose lepton or an isolated track respectively. Lepton $p_T$ ($E_T$) thresholds are documented in Table~\ref{tab:cats}.

Several standard model processes can mimic the trilepton signature with three leptons or two leptons and one isolated track. We remove backgrounds with on-shell Z boson by requiring that the invariant mass of the two opposite charged lepton-lepton or lepton-track pairs be between $76$ and $106\,\mathrm{GeV}/c^2$ and above a minimum invariant mass of $20$, $13\,\mathrm{GeV}$ to suppress Drell-Yan background and background from $\rm J/\Psi$ and $\Upsilon$ resonances. Additionally we demand $\slashed{E}_T>15\,\mathrm{GeV}$ and for the number of jets $N_{\mbox{Jets}}\leq 1$. The details of this analysis are documented in \cite{hep-ex, thesis, sthesis}.
\begin{table}[htbp]
\begin{minipage}{0.3\linewidth}
\centering
 \caption{\label{tab:cats} The $E_{\rm T}$ ($p_{\rm T}$) thresholds for electrons (muons, isolated tracks) for the five channels. $l_t$=tight lepton, $l_l$=loose lepton, and $T$=isolated track (lepton=$e,\mu$). }
 \begin{tabular}{cccc}
\hline 
\hline
Channel& && $E_{\rm T}$ ($p_{\rm T}$) {\small GeV (GeV/c)}\\ \hline
$\ttt$ & &&  15, 5, 5 \\
$\ttl$ & &&  15, 5, 10 \\
$\tll$ & &&  20, 8, 5 (10 if $\mu$) \\
$\ttT$ & && 15, 5, 5 \\
$\tlT$ & &&  20, 8 (10 if $\mu$), 5 \\
\hline
\hline
 \end{tabular}
\end{minipage}
\hspace{0.5cm}
\begin{minipage}{0.65\linewidth}
 \centering
\caption{\label{tab:bg} The number of expected events from background sources and for the mSUGRA point $m_0=60\,\mathrm{GeV/c^2}$, $m_{1/2}=190\,\mathrm{GeV/c^2}$, $\tan\beta=3$, $A_0=0$, $\mu>0$. The number of observed events in data in each channel is also shown. Uncertainties are statistical and systematic.
$l_t$=tight lepton, $l_l$=loose lepton, and $T$=isolated track (lepton=$e,\mu$). }
\vspace{1mm}
\begin{tabular}{ccccccc|c} 
\hline
\hline
Channel&&            $\ttt$&           $\ttl$&          $\tll$&           $\ttT$&           $\tlT$&       $\Sigma$channels\\ 
\hline
Drell-Yan&&   0.05&          0.01&          0.0&           1.63&          1.32&           \\
Diboson&&        0.29&          0.20&          0.08&           0.61&          0.38&           \\
Top-pair&&  0.02&          0.01&         0.03&           0.22&          0.18&           \\
Fake lepton&& 0.12&          0.04&         0.03&           0.75&          0.41&           \\
\hline
Total &&0.49&          0.25&          0.14&          3.22&          2.28&       6.4\\
Uncertainty&& $\pm$0.09&  $\pm$0.04&   $\pm$0.03&      $\pm$0.72&     $\pm$0.63&  \\
\hline
Observed&&     1&             0&            0&              4&            2&         7\\
\hline
SUSY Signal&&  2.3&          1.6&          0.7&            4.4&          2.4&        11.4\\
\hline
\hline
\end{tabular}
\end{minipage}
\end{table}

The leptonic decays of $WZ$, $ZZ$ and $t\bar{t}$ can produce three or more leptons. Dilepton processes, such as Drell-Yan or $WW$ accompanied by a bremstrahlung photon conversion, a fake lepton or an isolated track are other sources of background. W production in association with jets results in significant background when one jet gives a fake lepton and another an isolated track.
We estimate $WZ$, $ZZ$, Drell-Yan (+ brem conversion) and $WW$ (+ brem conversion) backgrounds using simulation. The rate for an additional isolated track is measured in data and applied to Monte Carlo events; the number of events with one fake lepton is measured in data. Background predictions are verified in control regions prior to revealing candidate trilepton events in data. The background expectations and the observed number of events in data are shown in Table~\ref{tab:bg}.

\section{Results}
\begin{figure*}[htbp]
\begin{minipage}{0.49\linewidth}
\centering
\includegraphics[width=\linewidth]{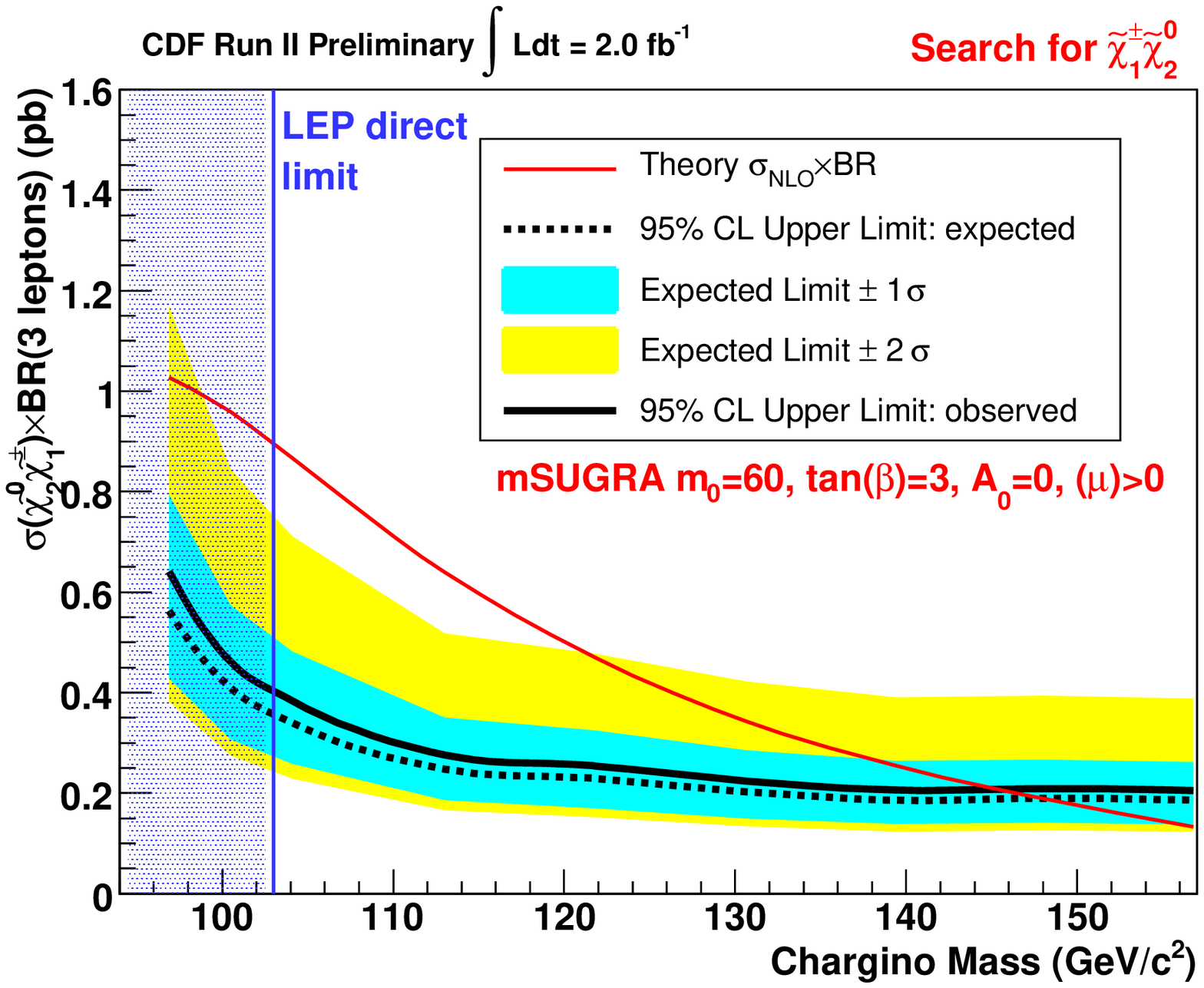}
(a)
\end{minipage}
\begin{minipage}{0.49\linewidth}
\centering
\includegraphics[width=\linewidth]{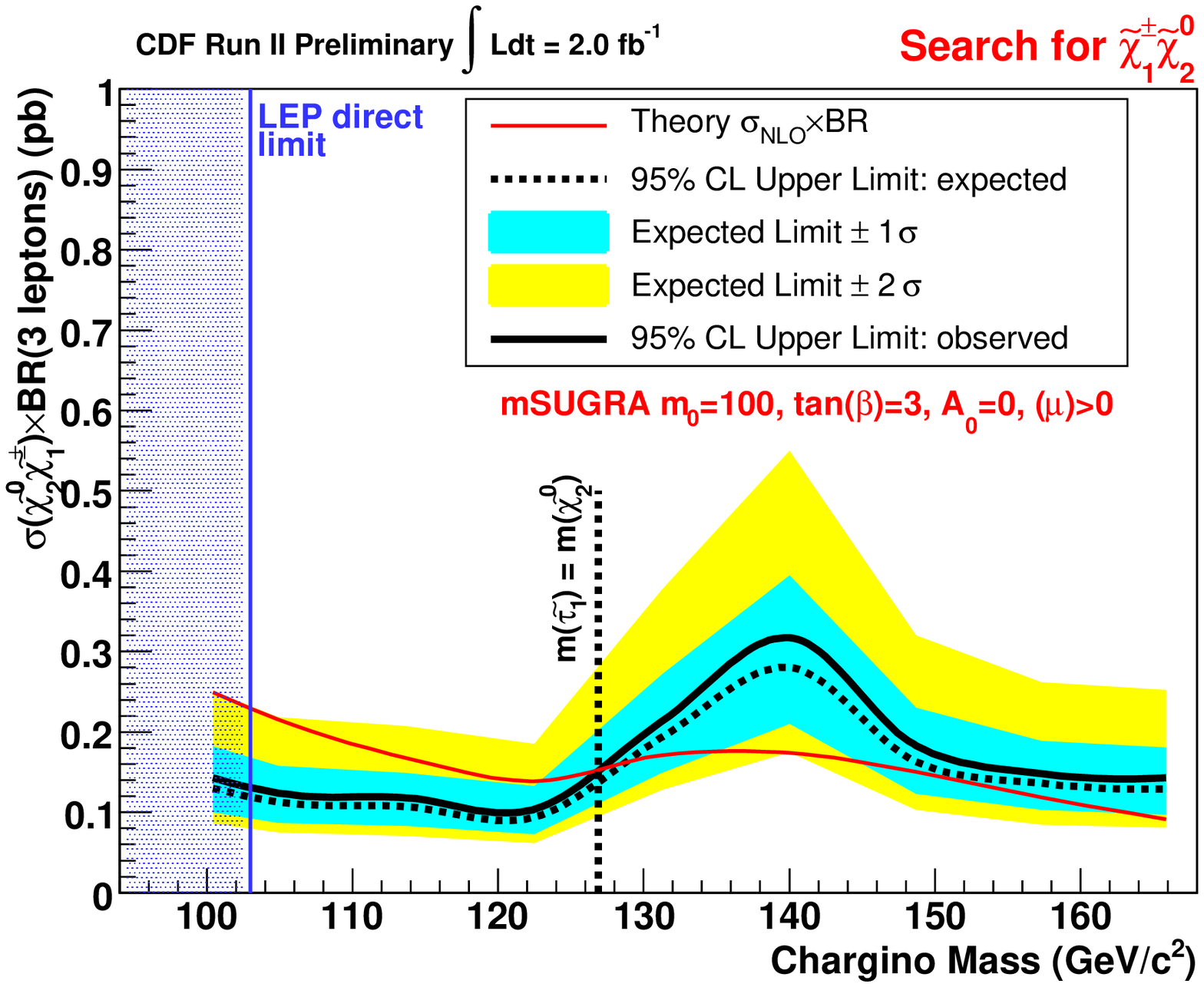}
(b)
\end{minipage}
\caption{Observed and expected $\sigma\times{\cal B}$ and prediction from theory as a function of chargino mass.
We rule out chargino masses below $145\,\mathrm{GeV/c^2}$ (where the theory and the experimental
curves intersect) for (a)~$m_0=60\,\mathrm{GeV/c^2}$ and below $127\,\mathrm{GeV/c^2}$ for (b)~$m_0=100\,\mathrm{GeV/c^2}$. \label{prl_fig3}} 
\end{figure*}

\begin{figure*}[htbp]
\begin{minipage}{0.48\linewidth}
\centering
\includegraphics[width=\linewidth]{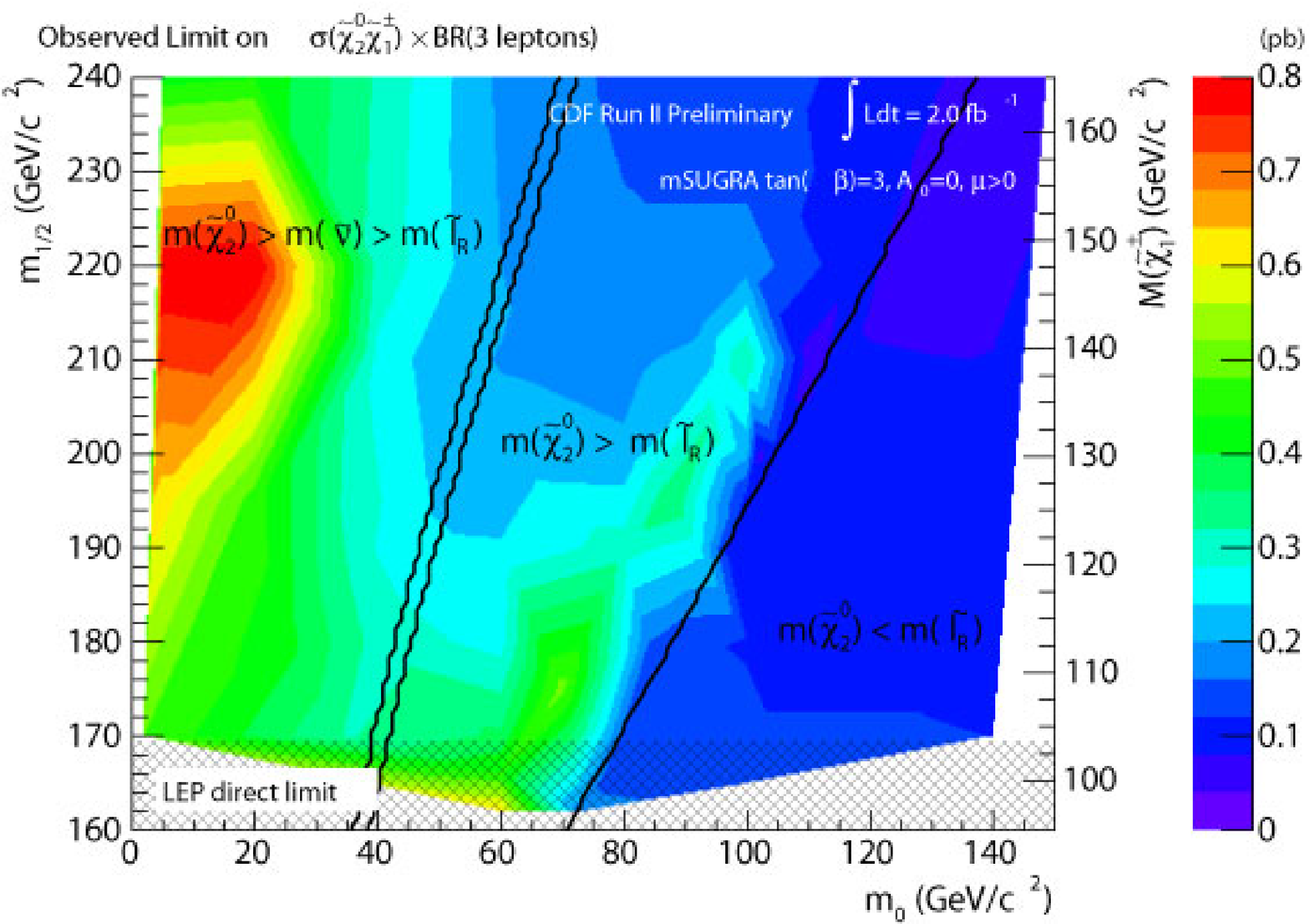}
\caption{Observed $95\%$ confidence level limit on $\sigma\times{\cal B}$ in mSUGRA with $\tan\beta=3$, $A_0=0$, $\mu>0$.} \label{observedlimit_regions}
\end{minipage}
\hspace{0.2cm}
\begin{minipage}{0.48\linewidth}
\centering
\includegraphics[width=\linewidth]{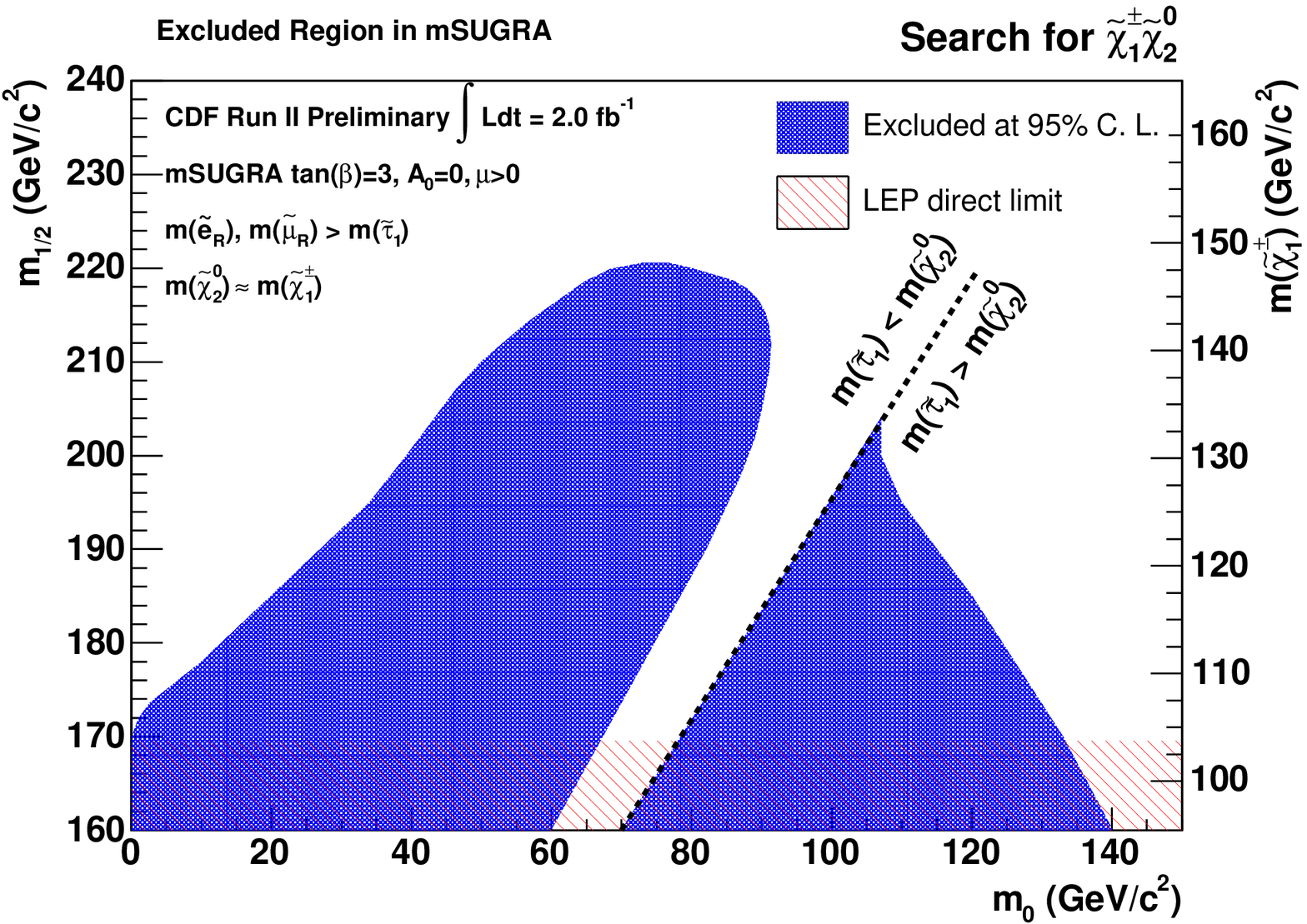}
\caption{Excluded regions in mSUGRA as a function of $m_0$ and $m_{1/2}$ for $\tan\beta=3$, $A_0=0$, $\mu>0$. Corresponding $\tilde{{\chi}}^\pm_{1}$ masses are shown on the right-hand axis.} \label{prl_fig2}
\end{minipage}
\end{figure*}

Table~\ref{tab:bg} shows that the observed number of events is consistent with the expected number of events from standard model background in each channel; no evidence for physics beyond the standard model is found. We calculate $95\%$ confidence level limits on the cross section for chargino-neutralino production multiplied by the branching ratio into three leptons. We obtain the acceptance from Monte Carlo samples for each set of mSUGRA parameters and the number of observed events and expected background events in Table~\ref{tab:bg}. The results are shown in Fig.~\ref{observedlimit_regions} as a function of the mSUGRA parameters $m_0$ and $m_{1/2}$; the other parameters are fixed to $\tan\beta=3$, $A_0=0$ and $\mu>0$. We compare the observed and expected $95\%$ C.L. limit and the theory expectation in a region where $m(\tilde{\tau}_1^\pm)<m(\tilde{\chi}_2^{0})$ for $m_0=60\,\mathrm{GeV/c^2}$ (Fig.~\ref{prl_fig3} (a)) and where $m(\tilde{\tau}_1^\pm)>m(\tilde{\chi}_2^{0})$ for $m_0=100\,\mathrm{GeV/c^2}$ (Fig.~\ref{prl_fig3} (b)) and derive limits on the chargino mass of $145\,\mathrm{GeV/c^2}$ and $127\,\mathrm{GeV/c^2}$. Fig.~\ref{prl_fig2} shows the limit on the chargino mass as a function of $m_0$. The exclusion is divided into two regions depending on whether the stau is heavier or lighter than the neutralino. For $m(\tilde{\tau}_1^\pm)<m(\tilde{\chi}_2^{0})$ the chargino and neutralino mainly decay into on-shell sleptons; for $m(\tilde{\tau}_1^\pm)>m(\tilde{\chi}_2^{0})$ the chargino and neutralino mainly decay via off-shell bosons or off-shell sleptons. The reach of this analysis is different for these regions since the branching ratio of $\tilde{\chi}_1^\pm\tilde{\chi}_2^0$ into three leptons has a strong dependence on the dominant decay channel\cite{hep-ph}. As the decay via sleptons always leads to leptons the highest limit on the chargino mass can be set in the region where this decay is dominant. If $m(\tilde{\tau}_1^\pm)\lessapprox m(\tilde{\chi}_2^{0})$, no exclusion is claimed: In the decay $\tilde{\chi}_2^0\to\tilde{l}^\pm l^\mp$ a lepton that is too soft to be detected is produced and the observed limit increases. This effect can also be seen in Figs.~\ref{observedlimit_regions} and~\ref{prl_fig3} (b). A more detailed study of the results of this analysis can be found in \cite{hep-ph}.

\end{document}